%% file: main.tex
\RequirePackage{snapshot}
\documentclass[sigconf]{acmart}
\renewcommand\footnotetextcopyrightpermission[1]{}
\usepackage{amsmath,amsfonts}
\usepackage{algorithmic}
\usepackage{graphicx}
\usepackage{textcomp}
\usepackage{xcolor}
\usepackage{subcaption}
\usepackage{cleveref}
\usepackage[inline]{enumitem}
\usepackage{todonotes}
\usepackage{comment}
\usepackage{booktabs}
\usepackage{url}
\usepackage[htt]{hyphenat}
\usepackage{textcomp}

\newcommand{\midtilde}{\raisebox{-0.25\baselineskip}{\textasciitilde}}

\begin{document}

\copyrightyear{2021}
\acmYear{2021}
\setcopyright{acmcopyright}\acmConference[UCC'21]{2021 IEEE/ACM 14th International Conference on Utility and Cloud Computing}{December 6--9, 2021}{Leicester, United Kingdom}
\acmBooktitle{2021 IEEE/ACM 14th International Conference on Utility and Cloud Computing (UCC'21), December 6--9, 2021, Leicester, United Kingdom}
\acmPrice{15.00}
\acmDOI{10.1145/3468737.3494104}
\acmISBN{978-1-4503-8564-0/21/12}

\title{
    Predictive Auto-scaling with OpenStack Monasca
}

\author{Giacomo Lanciano}
\affiliation{
    \institution{Scuola Normale Superiore}
    \city{Pisa}
    \country{Italy}
}
\email{giacomo.lanciano@sns.it}
\orcid{0000-0002-7431-8041}

\author{Filippo Galli}
\affiliation{
    \institution{Scuola Normale Superiore}
    \city{Pisa}
    \country{Italy}
}
\email{filippo.galli@sns.it}
\orcid{0000-0002-2279-3545}

\author{Tommaso Cucinotta}
\affiliation{
    \institution{Scuola Superiore Sant'Anna}
    \city{Pisa}
    \country{Italy}
}
\email{tommaso.cucinotta@santannapisa.it}
\orcid{0000-0002-0362-0657}

\author{Davide Bacciu}
\affiliation{
    \institution{University of Pisa}
    \city{Pisa}
    \country{Italy}
}
\email{davide.bacciu@di.unipi.it}
\orcid{0000-0001-5213-2468}

\author{Andrea Passarella}
\affiliation{
    \institution{National Research Council}
    \city{Pisa}
    \country{Italy}
}
\email{andrea.passarella@iit.cnr.it}
\orcid{0000-0002-1694-612X}

\begin{abstract}
    \input{abstract}
\end{abstract}

\keywords{Elasticity auto-scaling, Time-series forecasting, Predictive operations, OpenStack}

\settopmatter{printfolios=true}
\maketitle
\pagestyle{empty}
\pagenumbering{gobble}  

\input{intro}
\input{relwork}
\input{background}
\input{approach}
\input{experiments}
\input{conclusions}

\bibliographystyle{ACM-Reference-Format}
\bibliography{biblio-nourl}

\end{document}

%% file: abstract.tex
Cloud auto-scaling mechanisms are typically based
on \textit{reactive} automation rules that scale a cluster whenever
some metric, e.g., the average CPU usage among instances, exceeds a
predefined threshold. Tuning these rules becomes particularly
cumbersome when scaling-up a cluster involves non-negligible times to
bootstrap new instances, as it happens frequently in production cloud
services.

To deal with this problem, we propose an architecture for auto-scaling cloud services
based on the status in which the system is expected to evolve in the near future.
Our approach leverages on time-series forecasting techniques, like
those based on machine learning and artificial neural networks, to
predict the future dynamics of key metrics, e.g., resource consumption metrics,
and apply a threshold-based scaling policy on them.
The result is a \textit{predictive} automation policy that is able, for instance,
to automatically anticipate peaks in the load of a cloud application and trigger
ahead of time appropriate scaling actions to accommodate the expected increase in
traffic.

We prototyped our approach as an open-source OpenStack component,
which relies on, and extends, the monitoring capabilities offered by Monasca, resulting
in the addition of predictive metrics that can be leveraged by orchestration components
like Heat or Senlin. We show experimental results using a recurrent neural network and
a multi-layer perceptron as predictor, which are compared with a simple linear regression
and a traditional non-predictive auto-scaling policy. However, the proposed framework allows
for the easy customization of the prediction policy as needed.

%% file: intro.tex
\section{Introduction}

Information and communications technologies have been undergoing a
steep change over the last decade, where the massive availability of
low-cost high-bandwidth connectivity has brought to a tremendous push
towards distributed computing paradigms. This has led to the raise of
cloud computing technologies~\cite{Buyya11}, where the developments in a number of
virtualization technologies, spanning across the computing, networking
and storage domains, has facilitated the decoupling between management
of the physical infrastructure from the cloud services provisioned on
top of them. In the last decade, cloud computing has evolved from
infrastructure-as-a-service (IaaS) provisioning scenarios, where
physical servers were simply migrated within virtual machines (VMs),
to the nowadays platform-as-a-service (PaaS) era in which
\emph{native} cloud applications and services are developed relying
heavily on a plethora of networking, storage, security,
load-balancing, monitoring and orchestration services (XaaS -
everything-as-a-service) made available within cloud infrastructures
and their automation capabilities~\cite{Brunner15}.

A key success factor of cloud computing relies on the ability of cloud
providers to manage the infrastructure 24/7, promptly addressing and
resolving any issue that may occur at run-time, including both
hardware faults and possible software malfunctioning. This is made
possible by the use of appropriate data-center designs (i.e.,
fault-independent zones, redundant powering and cooling
infrastructures and multi-path networking topologies) coupled with the
use of feature-rich resource managers and orchestrators.

A \emph{cloud orchestrator} is the software at the center of that
concept of ``rapid provisioning'' with ``minimal management effort or service provider interaction'' from the quite popular cloud computing
definition from NIST~\cite{NISTDefCC11}. It includes key automation features that
give a cloud infrastructure self-healing and self-management
capabilities, thanks to the deployment of fine-grained monitoring
infrastructures that allow for triggering a number of automation
rules, to deal with a plethora of different issues: from handling
automatically hardware faults (individual computing, networking or
storage elements failing), to the automatic resizing of elastic and
horizontally scalable services, in order to cope with dynamic traffic
conditions and mitigate possible overloading conditions. A quite
popular cloud orchestrator is for example the open-source OpenStack
software.

The ``elasticity'' of cloud services, namely their capability to
expand their deployment over more and more instances (VMs, containers
or physical services if allowed) as the workload increases, allows for
keeping a stable performance (response times observed by individual
clients) despite fluctuations of the overall aggregated traffic
towards the services. This is realized through the use of control
loops, where an elasticity controller may decide to instruct scale-out
or scale-in operations, based on monitoring observations made over
time for individual services. These decisions are typically made on
the basis of monitored infrastructure-level resource consumption
metrics (e.g., CPU, networking, storage load), as well as
application-level metrics (e.g., response times, connection errors,
timeouts).

Traditionally, elasticity loops are performed applying a number of
\emph{threshold-based} rules: when a given metric (or a combination of
metrics) exceeds some warning or critical threshold, a scale-out
operation is triggered (e.g., resource utilization beyond certain
limits or service latency beyond acceptable values). Also, in order to
avoid continuous scaling operations due to the inherent fluctuations
of the monitored metric(s), it is typical to have a
high-water/low-water double-threshold set-up, where a scale-up is
performed when the monitored metric grows beyond a high threshold,
whilst a scale-down is performed when it returns below a much lower
threshold. Moreover, in order to avoid scale-up operations on
transient metric peaks, it is commonplace to require a violation of
the threshold for a few consecutive observations (typically, 3 samples
or 5 samples), before triggering the action.

Designing such seemingly simple control loops is accompanied by a set of
challenges, such as: determining the most effective key performance indicators
(KPIs) and the frequency at which they should be monitored; tuning the scaling
policy so that it can quickly adapt to substantial changes in the application
workloads, while being robust to fluctuations; estimating how many and which
type of resources are required in order to handle the new conditions; deciding
which scaling strategy (e.g., horizontal or vertical) is best suited.

\begin{figure}[t]
  \includegraphics[width=0.9\columnwidth,trim=10 0 0 10, clip]{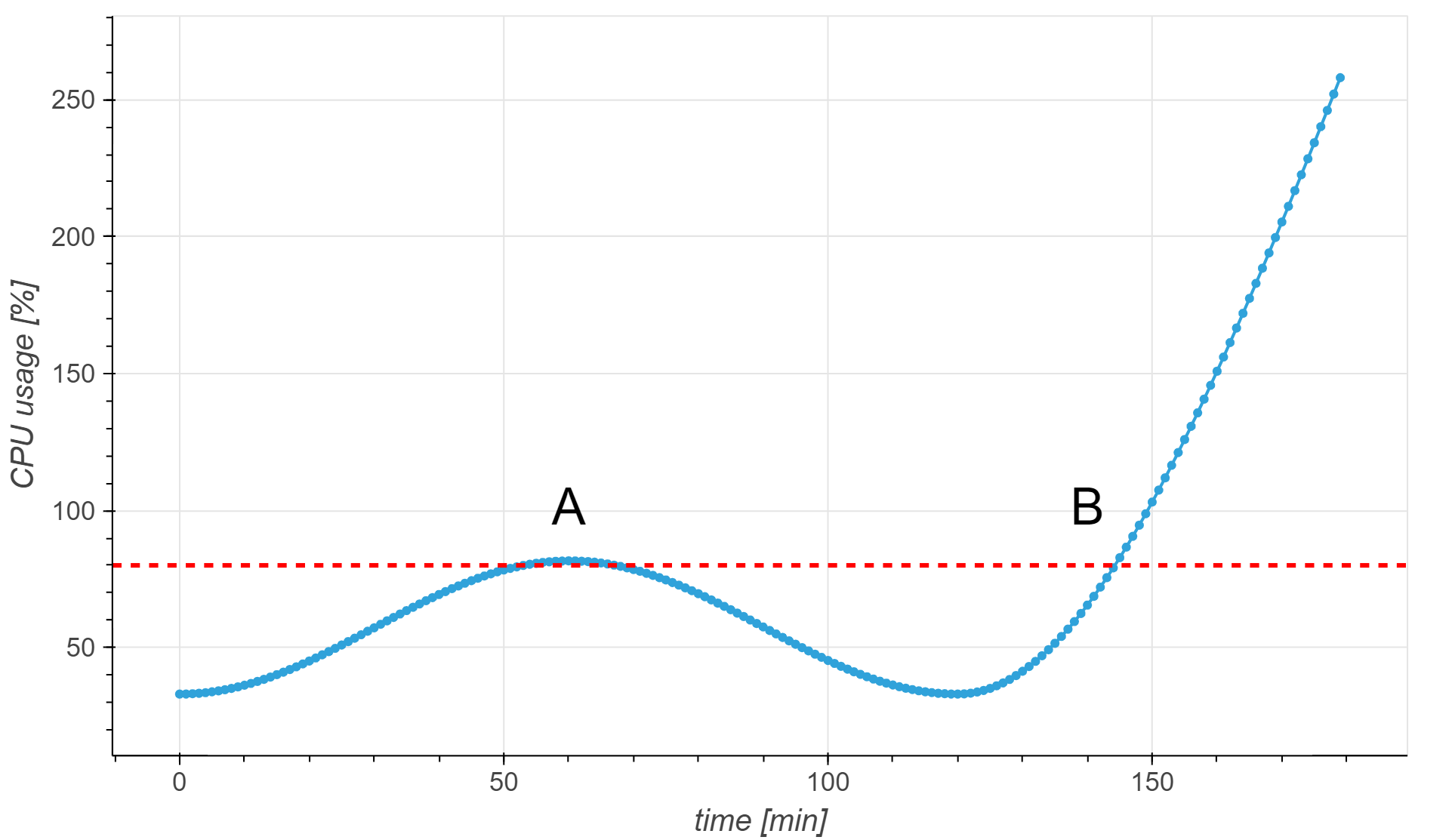}
  \caption{Example of load profile.}
  \label{fig:load-examples}
\end{figure}

Despite the mentioned precautions, control loops remain fundamentally ``dumb'',
in the sense that they do not account for the rich dynamic of the observed
metrics. As a simple example, consider the conceptual metric evolution depicted
in \Cref{fig:load-examples}. Here, we can see two scenarios where a metric
starts growing at quite different rates. A threshold-based scale-out would treat
these two conditions in pretty much the same way. However, a human operator
observing these curves, understands quite easily that scenario B needs a quicker
and more urgent action than scenario A. Although, this depends also exactly on the
time needed to bring up a new instance.

In the depicted scenario, a fundamental role may be played by
metric prediction and forecasting techniques, which, if properly
integrated within the automation rules of a cloud orchestrator,
may constitute a powerful tool to embed more intelligence within
elasticity controllers.

\subsection{Contributions}

This paper proposes an architecture for integrating predictive
analytics within an OpenStack cloud orchestration engine.  The
contributions of this paper include: 1) a general architecture for
performing \emph{predictive operations} on a cloud infrastructure,
based on smart metric forecasts that can be obtained via machine learning (ML) or
artificial intelligence (AI); 2) an open-source implementation of the
metric forecast component within OpenStack, leveraging on the Monasca
monitoring architecture, that automatically computes metric forecasts
making them available as additional metrics that can be leveraged
within the system at any level; 3) a few simple metric predictors
implemented in the framework as customizable predictors based on
linear regression (LR), multi-layer perceptrons (MLPs) and recurrent neural
networks (RNNs); the proposed architecture is expandable, so it eases the
implementation of additional predictive models in the code; 4) results
from real experimentation on a simple use-case based on synthetic
workload where we exploit the proposed architecture to create
\emph{predictive elasticity rules} using Monasca and Senlin,
exploiting the native capabilities of OpenStack so to realize
intelligent automation rules.

%% file: relwork.tex
\section{Related Work}

ML and AI techniques have been
investigated for a number of tasks related to resource management in
cloud computing infrastructures, in the context of elasticity for both
general public cloud services and private cloud infrastructures.  In
the following, we provide an overview of key research papers dealing
with intelligent elasticity management based on metric forecasting for
general public cloud services. Then we present related research in
predictive autoscaling for private cloud infrastructures with a focus
on network function virtualization (NFV)~\cite{NFV} and deployment of
distributed elastic service chains. Finally, we include a short review
of elasticity control solutions based on learning methods, such as
reinforcement learning (RL).

\subsection{Predictive elasticity in cloud computing}

In~\cite{Borkowski16-UCC}, a multi-layer neural network has been used to
predict the resource usage of tasks performing continuous integration
of several repositories from the Travis openly available
data. Using a per-repository trained model, an accuracy at least $20\%$ and
up to $89\%$ better than a baseline linear regression has been
achieved.

In~\cite{Roy11-CLOUD}, model-predictive techniques have been used to
track and predict the workload variability so to optimize resource
allocation in elastic cloud applications. The proposed technique
leverages an ARMA model for workload prediction, and it tries
to balance the advantages arising from dynamic elasticity, to the
cost due to applying the scaling decisions and reconfiguring
the cluster at each control period.

More recently, in the RScale framework~\cite{Kang20-UCC}, Gaussian
Process Regression, a probabilistic machine-learning method, has been
used to predict end-to-end tail-latency of distributed micro-services
workflows with generic DAG-alike topologies. This was evaluated on a
NSF Chamaleon test-bed, achieving similar accuracy but a smaller
predicted uncertainty with respect to using neural networks, but with
greatly enhanced execution-time overheads, and greater capability to
adapt on-line to dynamically changing workload/interference conditions.


In~\cite{Islam12-FGCS}, the problem of non-instantaneous instance
provisioning when using elastic scaling in cloud environments is
tackled, by proposing a predictive scaling strategy based on a
resource prediction model using neural networks and linear
regression. The method has been applied on an e-commerce application
scenario emulated through the well-established TPC-W~\cite{tpc2021} workload
generator and benchmarking application, deployed within AWS EC2.
Predictions made via neural networks enhance the accuracy by reducing
the mean average percentage error (MAPE) by roughly $50\%$ compared to linear
regression.


In~\cite{Bashar13-CloudNet}, Bayesian Networks are used in a
predictive framework to support automatic scaling decisions in cloud
services. However, the method is evaluated on synthetic applications
with exponential start times, duration and workload inter-arrival
patterns.


Other approaches exist that prefer to focus on real-time dynamic resource
allocation based on instantaneous monitoring rather than resource
estimations/predictions made ahead of time.  For example,
in~\cite{Nicodemus20-UCC}, a vertical elasticity management of containers has
been proposed, to adapt dynamically the memory allocated to containers in a
Kubernetes cluster, so to better handle the coexistence of containers with
heterogeneous quality of service (QoS) requirements (guaranteed, burstable and
best-effort). However, approaches of this type fall within the research
literature on classical reactive elasticity loops, which we omit for the sake of
brevity.

\subsection{Predictive elasticity for NFV services}

Predictive techniques in private cloud computing scenarios have also been
investigated in the context of NFV and Software Defined Networks (SDN) for
adapting available virtualized resources to varying
loads~\cite{Makaya15,Miyazawa15,Niwa15}.
This way, operators can benefit from proactive automation mechanisms to ensure
appropriate QoS for their cloud-native \textit{service-chains}.
Although, several challenges must be tackled in order to obtain an effective
solution to such problem. For instance:
\begin{enumerate*}[label=(\roman*)]
    \item correctly assessing which components need to be scaled to remove
    bottlenecks;
    \item optimizing the consolidation of virtual resources on the physical
    infrastructure;
    \item designing effective predictive models that prevent virtual resources
    from being under- or over-provisioned.
\end{enumerate*}~\cite{Fei2018}.

The authors of~\cite{rahman2018auto} describe the usage of \emph{ensembling}
techniques (i.e., combining the outputs from several models) for auto-scaling
purposes.
In~\cite{zaman2019novel}, instead, the authors propose a different approach
based on Long Short-Term Memory (LSTM) networks, a particular flavor of RNNs,
that is used for virtual network function (VNF) demand forecasting.
%
Nowadays, RNNs have been proven to be a powerful tool for time-series analysis,
being them forecasting~\cite{Rangapuram2018,Laptev2017,Cucinotta2021} or
classification~\cite{IsmailFawaz2019,Malhotra2017} tasks.
In particular, the \textit{sequence-to-sequence} architectural pattern, widely
adopted for machine translation and Natural Language Processing (NLP) tasks,
yields surprisingly good results~\cite{sutskever2014sequence}. Such
architectures typically consist of two distinct modules, the \textit{encoder}
and the \textit{decoder}.
In the context of NFV, sequence-to-sequence models can be used to capture the
complex relationships between VNF metric sequences and infrastructure metric
sequences~\cite{Cucinotta2021}. Notice that, since the information exchange
between the encoder and the decoder is restricted to the hidden state values,
one can even design a model that uses different metrics for inputs and outputs.

Forecasting accuracy can also be boosted by additional information about the
topology of the deployed VNFs, e.g., graph-like diagrams depicting the
interactions among the VMs belonging to the same VNFs~\cite{Miyazawa15,Niwa15}.
For instance, the authors of~\cite{Mijumbi17} propose a \emph{topology-aware}
time-series forecasting approach leveraging on graph neural networks
(GNNs)~\cite{Bacciu20}.

\subsection{Elasticity control with Reinforcement Learning}

The most widely used approach to elasticity control is \emph{static thresholding}
~\cite{carella2016extensible}. Even though it is a straightforward
heuristic, it can yield surprisingly good results when dealing with simple
systems.
However, in general, threshold policies require careful tuning for each service
whose elasticity must be controlled, making a generic approach impossible to
be adopted in practice (as it would eventually lead to over- or
under-provisioning).
To overcome such inconveniences, several \emph{dynamic thresholding} mechanisms
have been proposed to adapt thresholds to the current conditions
of the system. Such methods can be implemented with AI-based techniques as well,
like RL~\cite{arteaga2017adaptive, Arabnejad2017}.
However, RL algorithms usually come with demanding computing requirements that
often limit their applicability in real infrastructures.
For instance, the authors of~\cite{tang2015efficient} propose a Q-learning-based
algorithm to be used in an actual telco system. However, the developed agent is
allowed to take several unexpected decisions before converging to the optimal
policy. This is clearly not desired when deploying the system in a production
environment.
On the other hand, the authors of~\cite{Xiao2019} propose a rather successful
RL-based approach to VNF service-chains deployment that jointly minimizes
operation costs and maximizes requests throughput, also taking into account
different QoS requirements.

%% file: background.tex
\section{Background}
\label{sec:background}

\begin{figure*}[pt]
  \includegraphics[width=0.9\textwidth]{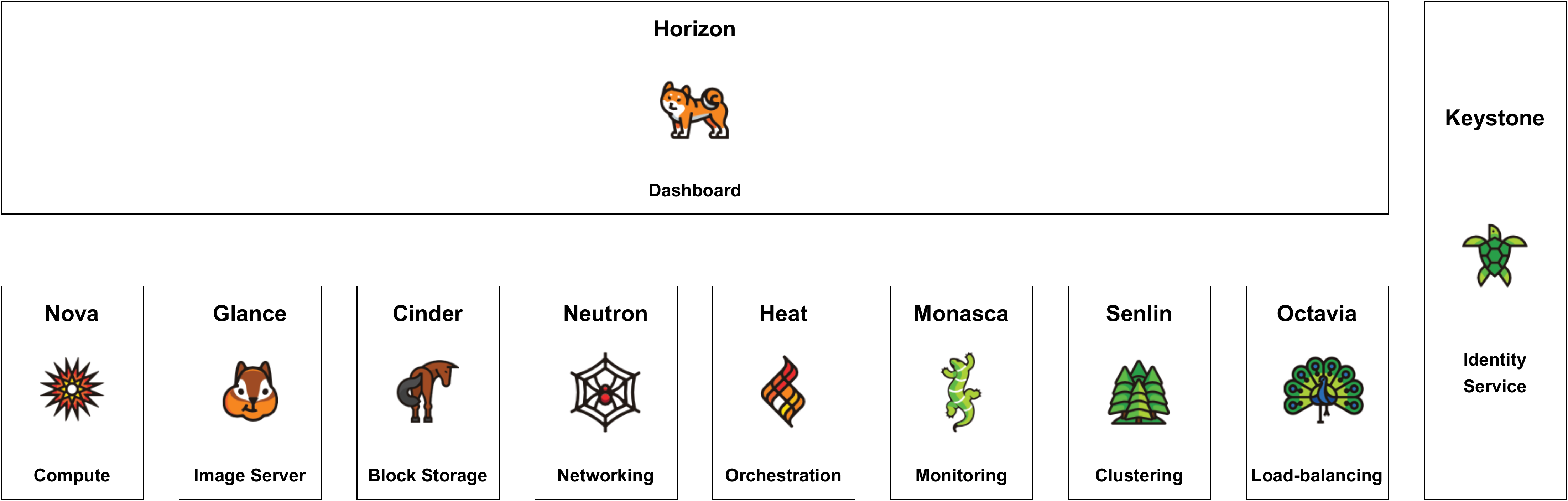}
  \caption{Overview of OpenStack key components.}
  \label{fig:openstack}
\end{figure*}

This section contains background concepts useful for a better understanding of
the approach proposed in \Cref{sec:approach}.
A number of key components of OpenStack architecture are detailed below, with
reference to \Cref{fig:openstack}.
Additionally, we include some well-known definitions around RNNs, which have
been used in our proposed architecture.

\subsection{Nova, Glance and Cinder}

Nova~\cite{OpenstackNova2021} is the OpenStack project providing the necessary
tools to provision and manage compute instances. It supports the creation of
VMs, bare metal servers and containers (limited). It
leverages on the Glance~\cite{OpenstackGlance2021} service for image management
and provisioning, and the Cinder~\cite{OpenstackCinder2021} service for
management of block storage that can be used to mount remote volumes in VM
instances.
The Nova architecture consists in a number of server processes, that communicate
each other via an RPC message passing mechanism, and a shared central database. The
core of the architecture is the \textit{compute} process, whose job is to manage
the underlying hypervisor exploiting the capabilities of libvirt.
The \textit{compute} process communicates with the Nova database
by proxying its queries via the \textit{conductor} process. The
\textit{scheduler} process instead is responsible for interfacing with the
compute instances placement service, whose behavior can be customized through
a number of plugins (filters).

\subsection{Neutron}

Neutron~\cite{OpenstackNeutron2021} is the OpenStack component in charge of
providing network-as-a-service connectivity among instances managed by Nova. It
allows for managing and customizing per-customer dedicated virtual networks with
their own numbering and DHCP configurations, and that can be enriched with
security capabilities including management of firewall rules and virtual private
networks (VPNs). Neutron used to include also load-balancing-as-a-service
capabilities, which have recently been engineered into the separate Octavia
service (see below).

\subsection{Monasca}

Monasca~\cite{OpenstackMonasca2021} is the OpenStack project
providing an advanced monitoring-as-a-service solution that is multi-tenant,
highly scalable, performant, and fault-tolerant.
It consists in a micro-service architecture, where each module is designed to
play a well-defined role in the overall solution (e.g., a streaming alarm engine
leveraging on Apache Storm, a notification engine, a persistence layer backed by
an efficient time-series database). The core of the architecture is a Kafka message
queue, that enables asynchronous communication among the components. Monasca
also includes an agent module, that is distributed on the machines hosting the
compute entities and is responsible for actually collecting the metrics and
forwarding them to the message queue through the REST APIs.

Apart from being usable in an OpenStack deployment, Monasca can also be deployed
in a Kubernetes environment as a standalone monitoring solution.

\subsection{Senlin}

Senlin~\cite{OpenstackSenlin2021} is the OpenStack project
providing the necessary tools to create and operate easily clusters of homogeneous
resources exposed by other OpenStack services.
The interactions between Senlin and the other OpenStack resources is enabled by
the \textit{profile} plugins. Once a profile type is chosen (e.g., a Nova
instance), a \textit{cluster} of resources which the profile refers to can be
created and associated with \textit{policies}. Such objects define how the
resources belonging to a cluster must be treated under specific conditions. For
instance, one can define a \textit{scaling} policy to automatically resize the
cluster, as well as a \textit{health} policy to replace the resources as needed,
or a \textit{load-balancing} policy to evenly distribute the workload.

Notice that, with respect to Heat's auto-scaling group abstraction, Senlin
provides a more fine-grained control over the management policies that should be
applied, as well as effective operation support tools. Indeed, for instance,
Senlin is used at large companies like Blizzard Entertainment to provide their
on-premise gaming servers with auto-scaling capabilities~\cite{Truong2019}.

\subsection{Octavia}

Octavia~\cite{OpenstackOctavia2021}, previously known as \textit{Neutron LBaaS},
is the OpenStack project providing a scalable and
highly-available load-balancing solution.
Load-balancing is fundamental feature for the cloud, as it enables a number of
other properties (e.g, elasticity, high-availability) that are considered of the
utmost importance for a modern production cloud environment.
Octavia delivers its services by managing a horizontally-scalable pool of Nova
instances (VMs, bare metal servers or containers), known as \textit{amphorae},
that leverage on the features provided by HAProxy. Such pool is orchestrated by
the \textit{controller}, that consists in a number of sub-components whose jobs
include handling API requests and ensuring the health of the amphorae.

\subsection{Recurrent Neural Networks}
\label{sec:rnn}

RNNs are commonly used for predicting uni-variate or multi-variate time-series
evolution in the future. In this paper, we used only uni-variate predictions.
Here, we recall basic concepts about how RNNs are trained and used, for the
sake of completeness.

As widely known~\cite{Goodfellow2016}, when predicting a target metric evolution using an RNN model,
future metric estimations are based on the current
metric value and an $H$-dimensional vector, computed recurrently from the past $I$
samples, representing its state.
The evolution of the model is then governed by the following equations:
\begin{align}
  s_{t} & = f_s(s_{t-1}, i_{t}) \\
  o_{t} & = f_o(s_{t}, i_{t})
\end{align}
where $f_s(s,i): \mathbb{R}^{H+I}\rightarrow \mathbb{R}^{H}$ acts on
$x,$ the concatenation of the hidden state vector $s \in \mathbb{R}^H$ and input
vector $i \in \mathbb{R}^I$, and $f_s =\text{ReLU}(W_{s}x + b_s)$,
with $W_s$ and $b_s$ being the weight and bias tensors
respectively, and $\text{ReLU}$ denoting the element-wise Rectified Linear Unit activation
function; $f_o: \mathbb{R}^{H+I}\rightarrow \mathbb{R}^{O}$ is the output
function defined as $\text{ReLU}(W_{o}x + b_o)$. Training of the learnable
parameter set $\theta = \{\theta_j\} =  \{W_s, W_o, b_s, b_o\}$ is achieved through the gradient
descent algorithm with momentum, so that at the $k$-th optimization step
the $j$-th parameter will be updated according to:
\begin{align}
  \mu_{j,k}      & = \beta \mu_{j,k-1} + \nabla J_{\theta_{j,k}}(D) \\
  \theta_{j,k+1} & =  \theta_{j, k} - \lambda\mu_{j,k}
\end{align}

With $\nabla J_{\theta_{j,k}}(D)$ being the gradient of the cost function
$J_{\theta_{j,k}}(D)$ computed with respect to parameter $\theta_j$ at
instant $k$ over
dataset $D$ of input-output pairs, $\lambda$ is the learning rate, and
$\beta$ is a hyper-parameter determining how much momentum $\mu_{j,k}$ is applied during the
gradient descent step, usually set to $0.9$.
Training continues until convergence, i.e. when the minimum value of the loss
function has been reached. A working definition of \textit{minimum value}
involves computing the loss function over a held-out validation  dataset $D'$ every
$K$  optimization steps, and continuing training as long as the validation
loss keeps decreasing. As the validation loss curves up again (as a
consequence of over-fitting) the optimal model is taken as the model corresponding to
the minimum of the validation loss.

%% file: approach.tex
\section{Proposed Approach}
\label{sec:approach}

\begin{figure}[t]
  \centering
  \includegraphics[keepaspectratio,width=0.99\columnwidth]{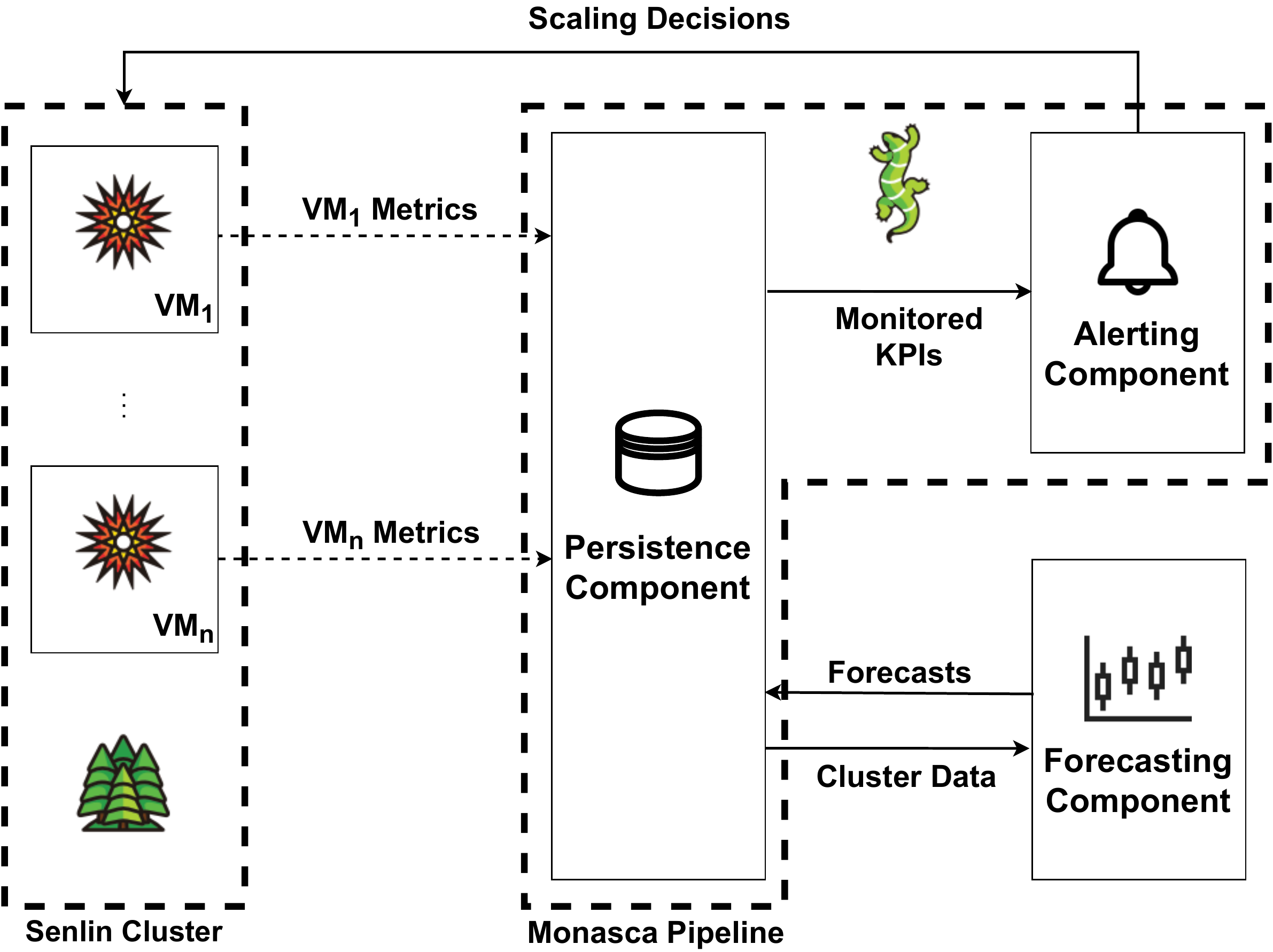}
  \caption{
    Architectural diagram of the proposed predictive auto-scaling approach.
  }
  \label{fig:arch-diagram}
\end{figure}

Our approach consists in a \textit{predictive} auto-scaling strategy, that
estimates the future state of the monitored system and takes scaling decisions
based on such estimates. Depending on the quality of the estimates, such a
strategy is able, for instance, to automatically anticipate peaks in the load of
a cloud application and trigger appropriate scaling actions to accommodate the
increasing traffic.
This feature allows for overcoming the main drawbacks of classical\textemdash
\textit{reactive}\textemdash auto-scaling approaches, that typically start
acting on the system when the load is already too high to be properly handled,
and the QoS is already degraded. Of course, operation teams could err on the
side of caution and tune the system such that the auto-scaling kicks in well in
advance with respect to the actual load peak. However, doing this entails paying
additional costs for provisioning resources that could be potentially not
needed. Resource over-provisioning is not a sustainable strategy in many
scenarios.
In the context of cloud applications, where certain components are typically
designed to be horizontally scaled to accommodate a varying level of traffic,
one should also take into account that spinning up new replicas can take a
non-negligible amount of time. One of the key benefits of using a predictive
approach is indeed the possibility to mitigate such a \textit{cold-start} effect,
by triggering the resource provisioning process at the right time for the
replica to be ready when the additional traffic starts hitting the service.

To demonstrate the effectiveness of our approach, and its relevance in the
context of modern cloud operations, we realized an implementation that is
designed to interoperate with an OpenStack deployment. In particular, our
implementation heavily relies on, and extends, the monitoring capabilities
offered by Monasca.
\Cref{fig:arch-diagram} shows how our predictive component enhances the typical
elasticity control loop of a horizontally-scalable set of Nova compute instances
(VMs). In our implementation, the actual resource orchestration is performed by
Senlin.

Our approach works as follows. As soon as new system-level measurements from
the compute instances are ingested and made available through the Monasca API, we
leverage on time-series forecasting techniques to predict the future dynamics of
some relevant infrastructure metrics. These are assumed to be key
indicators for the amount of load that the system is handling (e.g., CPU
utilization).
Notice that, depending on the chosen time-series forecasting algorithm, it might
be necessary to train the resulting model on a large portion of historical
monitoring data. In addition, given the high velocity at which the operating
conditions change, as typical in cloud environments, such a model most
likely has to be updated frequently to overcome the effects of concept drift
and keep getting accurate forecasts. At the moment, these aspects are assumed to
be handled offline, such that the forecasting component must be provided with a
pre-trained version of the chosen model that is ready to be used for inference.

The predictor is fed with a time-series in input that is obtained by aggregating
the time-series of the target metric as measured from the VMs composing the
elasticity group. The forecasts output by the predictor are then persisted back to Monasca,
such that other infrastructure components can consume them. In addition, these predicted metrics
become available to operators through standard monitoring dashboards (such as Grafana).
In our case, the predicted metrics
are used to feed a \textit{threshold-based} scaling policy. This type of
policies typically triggers a scale-out/scale-in action as soon as the specified
upper/lower threshold is repeatedly reached for a certain number of subsequent
observations. In our implementation, the threshold checks are performed by the
highly-scalable alerting pipeline provided by Monasca through Apache Storm.
Then, depending on the conditions of the system, Monasca notifies the Senlin
orchestration engine regarding the scaling actions to be performed.

As detailed in the next section, in our experimentation we considered as
predictors: a linear regressor, an MLP, and an RNN (see \Cref{sec:rnn}).

%% file: experiments.tex
\section{Experiments}
\label{sec:experiments}

In this section, we report experimental results showing the performance of the
approach described in \Cref{sec:approach}, compared to a classical
\textit{reactive} threshold-based scaling strategy.

\subsection{Experimental Set-up}

From an infrastructural standpoint, as explained in details in
\Cref{sec:approach}, our test application leveraged on: \textit{(i)} Senlin to
orchestrate a horizontally-scalable cluster of Nova instances; \textit{(ii)}
Octavia to provide the cluster with load-balancing capabilities; \textit{(iii)}
Monasca to ingest the system-level metrics and to trigger the scaling actions;
\textit{(iv)} the forecasting component, developed by us, to estimate the future
state of the system and to enable the \textit{proactive} auto-scaling strategy.
The Senlin cluster was configured to have a minimum of 2 active Nova instances
and to expand up to 5. Each instance was configured to run Ubuntu 20.04 cloud
image and to have 1 vCPU and 2 GB of RAM available.
The Octavia load-balancer was configured to distribute the traffic among the
active instances according to a simple \textit{round-robin} strategy.
Monasca was configured such that new CPU usage measurements were collected each
minute.
The forecasting component was configured to output a new prediction with the
same interval, using the last 20 minutes worth of data as input. In particular,
the input to the underlying forecasting model consisted in a time-series
reporting the sum of the CPU usage measurements of the currently active
instances. The output of the model was the estimated value in 15 minutes of the
same time-series. The output was then divided by the number of currently active
instances to get an estimate of the \textit{average CPU usage} in 15 minutes
(assuming the cluster size to be constant) and then persisted back to Monasca.
We used PyTorch to implement standard flavors of MLP and RNN (see
\Cref{sec:rnn}), while we used the implementation of the linear regressor
provided by Scikit-learn.

To implement the \textit{predictive} scaling strategy, the alerting component of
Monasca was configured to trigger a scale-out action whenever the
\textit{predicted} average CPU usage of the cluster, as outputted by the
forecasting component, reached the 80\% for 3 times in a row. On the other hand,
to implement the classical \textit{reactive} scaling strategy, the alerting
component was set to track the \textit{actual} average CPU usage of the cluster
in a similar fashion.
In both cases, the alerting component was set to trigger a scale-in action
whenever the \textit{actual} average CPU usage of the cluster reached the 15\%
for 3 times in a row. Notice that such settings impose a delay of at least 3
minutes for an action to be triggered. In addition, each scaling action could
adjust the size of the cluster by 1 instance only and could only take effect if
it was triggered after at least 20 minutes since the last effective action
(i.e., \textit{cooldown} period).
After a scale-out action was triggered and the resulting new instance was
spawned, we imposed an additional delay of 6 minutes before such instance could
start serving requests, emulating what could happen in a real production service
deployment, where spinning up a new VM hosting complex software might take a
non-negligible amount of time.

During our test runs, we generated traffic on the system using
\textit{distwalk}~\cite{CucinottaDistwalk2021}, an open-source distributed
processing emulation tool developed by us. Such tool consists in a server
module, that is started at boot on each instance and waits for TCP requests, and
a client module, that sends, in this case, requests to the load-balancer with
the aim of increasing the CPU utilization of the instances.
The client was configured such that it spawned 6 threads, each one provided with
a \midtilde1.5h-long trace reporting the operation rates (i.e., requests per
second) that should be maintained for an interval of one minute each. Each
thread was also forced to break its work in 1000 sessions, such that a new
session opened a new connection with the load-balancer, that in turn selected a
(possibly different) target instance.

The experiments were carried out using an all-in-one deployment of OpenStack
(\textit{Victoria} release), that was deployed using the tools provided
by~\cite{OpenstackKolla2021}. Notice that, in this case, the various OpenStack
services are installed and operated within several Docker containers, not on
bare metal.
%
The deployment was hosted on a Dell R630 dual-socket test-bed, equipped with:
2 Intel Xeon E5-2640 v4 CPUs (2.40 GHz, 20 virtual cores each); 64 GB of RAM;
Ubuntu 20.04.2 LTS operating system.

The implementation of the forecasting component was written in Python and is
released under an open-source license~\cite{Lanciano2021MonascaPredictor}.
For the sake of reproducibility, this work comes with a publicly available
companion repository~\cite{Lanciano2021Companion} including: the Heat templates
used to set up the infrastructure; the configuration files for the tools we used
to run our experiments; the raw results produced by our experiments; the code
required to generate the results (i.e.,
\Cref{fig:results,tab:percentiles,tab:overhead}) included in this paper; the
synthetic dataset used to train the neural models and their pre-trained
versions; etc.

\subsection{Results}

\begin{figure*}[p!]
    \centering
    \subcaptionbox{
        Static scaling policy - CPU usage\label{fig:run-static}}{
        \includegraphics[keepaspectratio,width=0.48\textwidth,trim=10 0 0 60, clip]{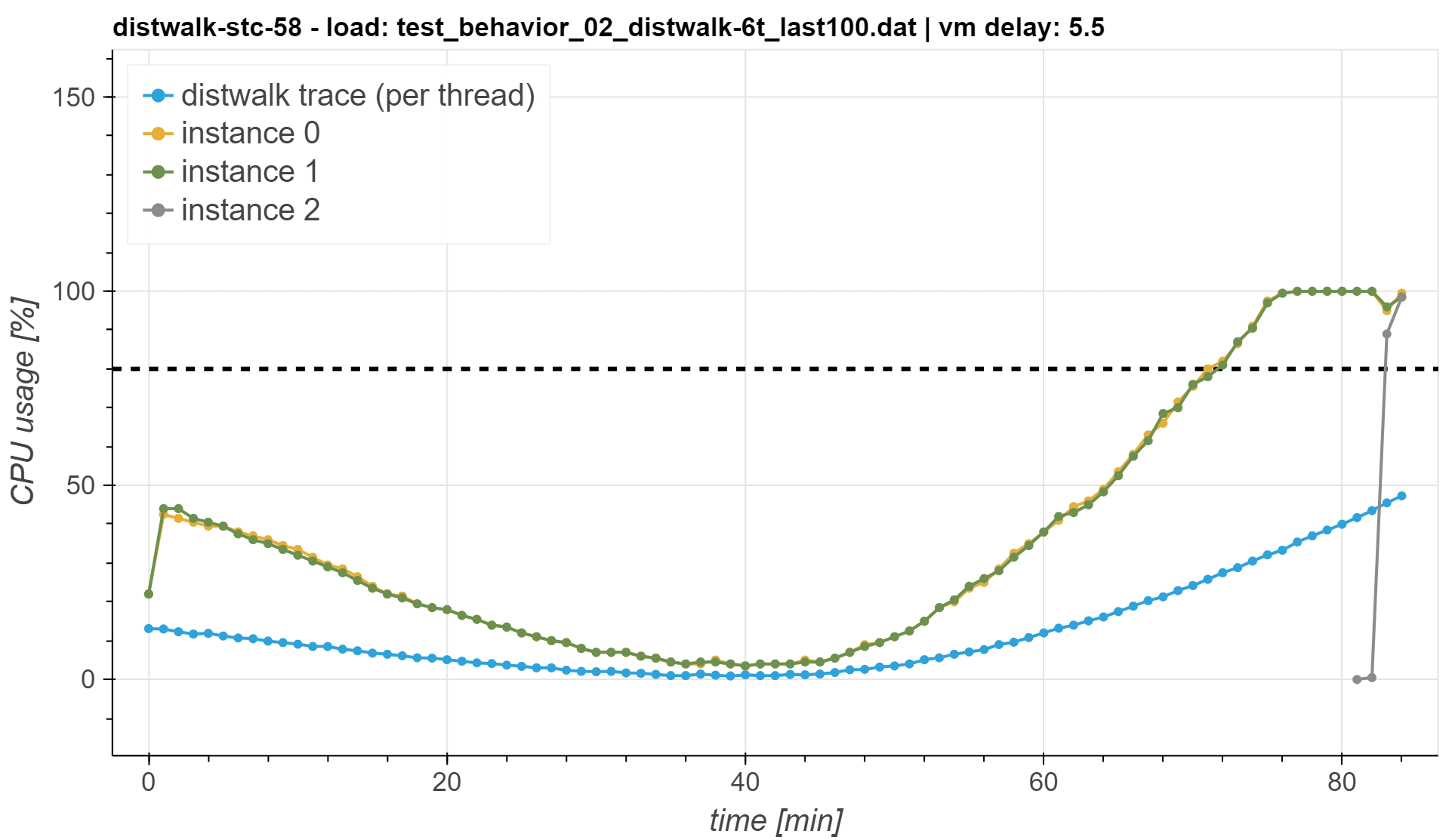}
    }
    \hfill
    \subcaptionbox{
        Static scaling policy - client-side response time\label{fig:run-static-times}}{
        \includegraphics[keepaspectratio,width=0.48\textwidth,trim=10 0 0 60, clip]{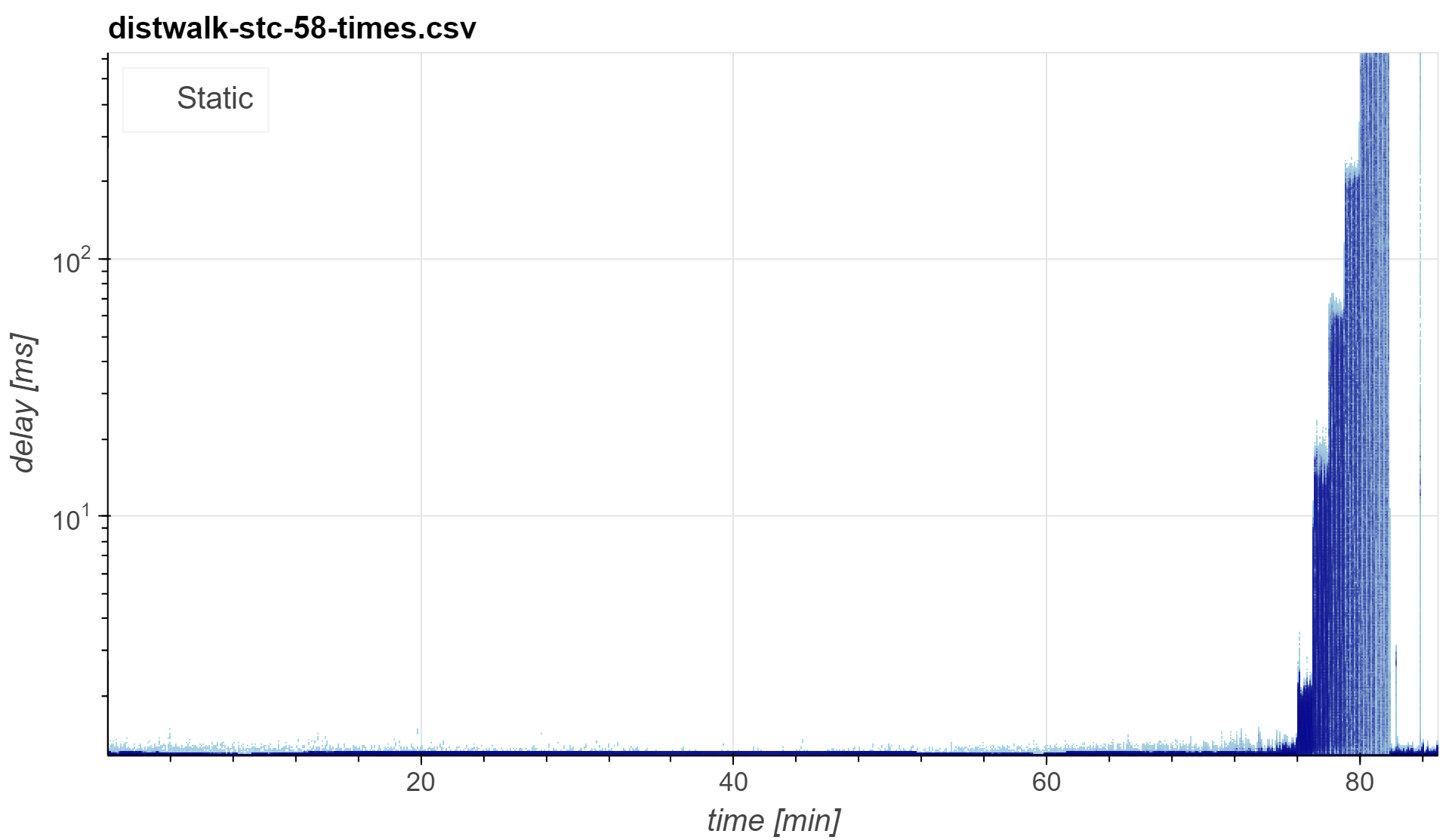}
    }
    \subcaptionbox{
        LR-based scaling policy - CPU usage\label{fig:run-lin}}{
        \includegraphics[keepaspectratio,width=0.48\textwidth,trim=10 0 0 60, clip]{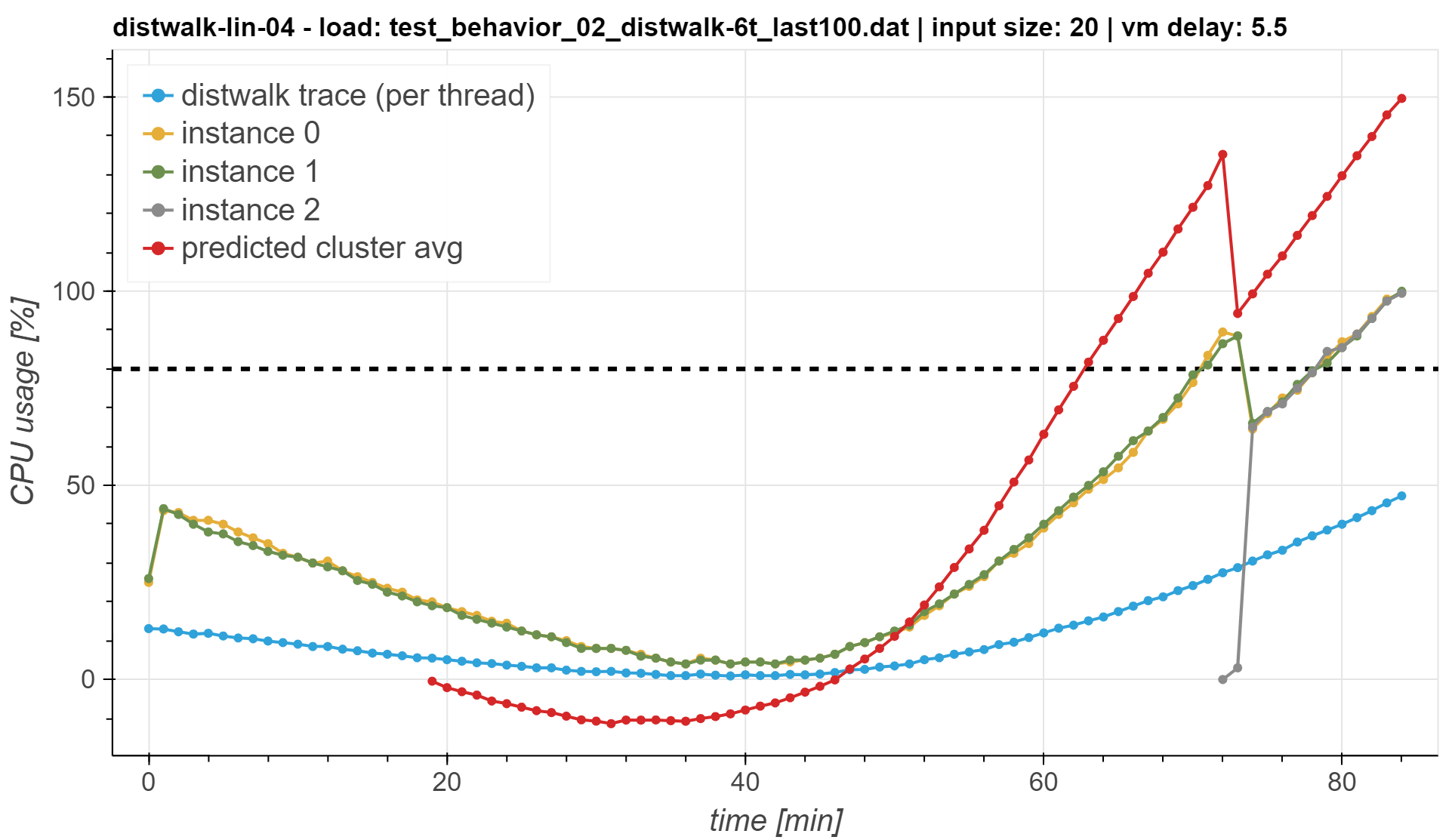}
    }
    \hfill
    \subcaptionbox{
        LR-based scaling policy - client-side response time\label{fig:run-lin-times}}{
        \includegraphics[keepaspectratio,width=0.48\textwidth,trim=10 0 0 60, clip]{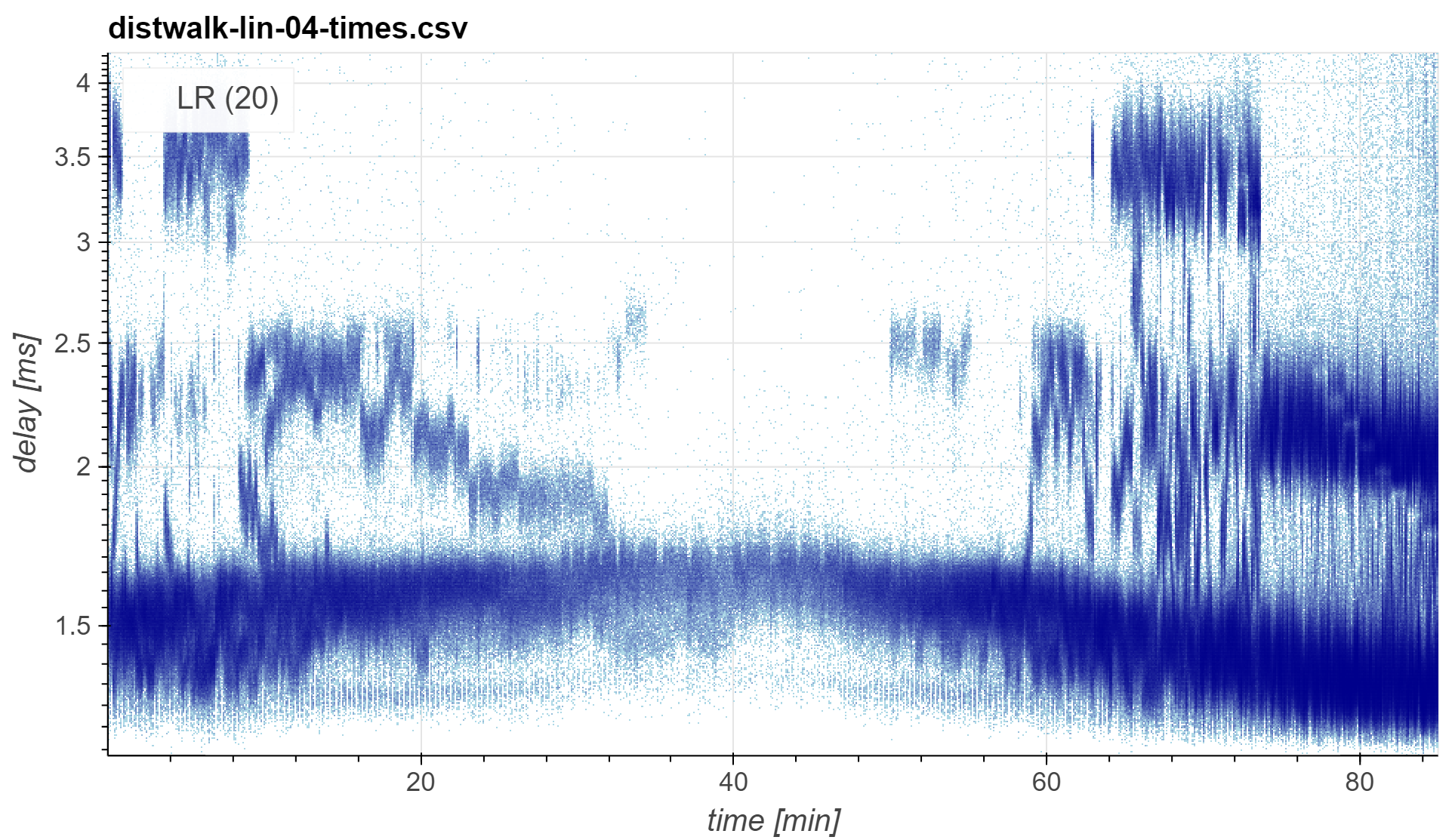}
    }
    \subcaptionbox{
        MLP-based scaling policy - CPU usage\label{fig:run-mlp}}{
        \includegraphics[keepaspectratio,width=0.48\textwidth,trim=10 0 0 60, clip]{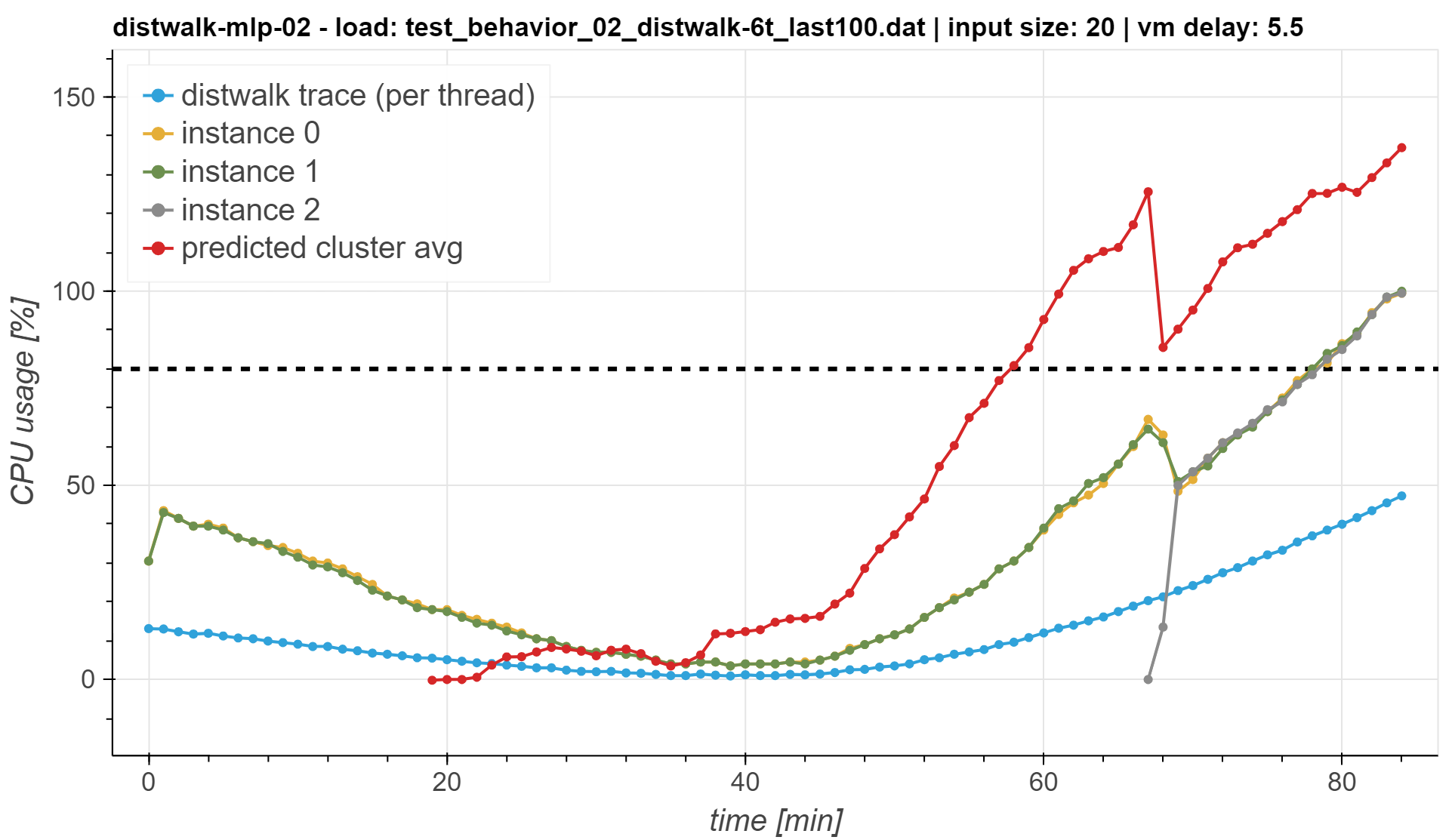}
    }
    \hfill
    \subcaptionbox{
        MLP-based scaling policy - client-side response time\label{fig:run-mlp-times}}{
        \includegraphics[keepaspectratio,width=0.48\textwidth,trim=10 0 0 60, clip]{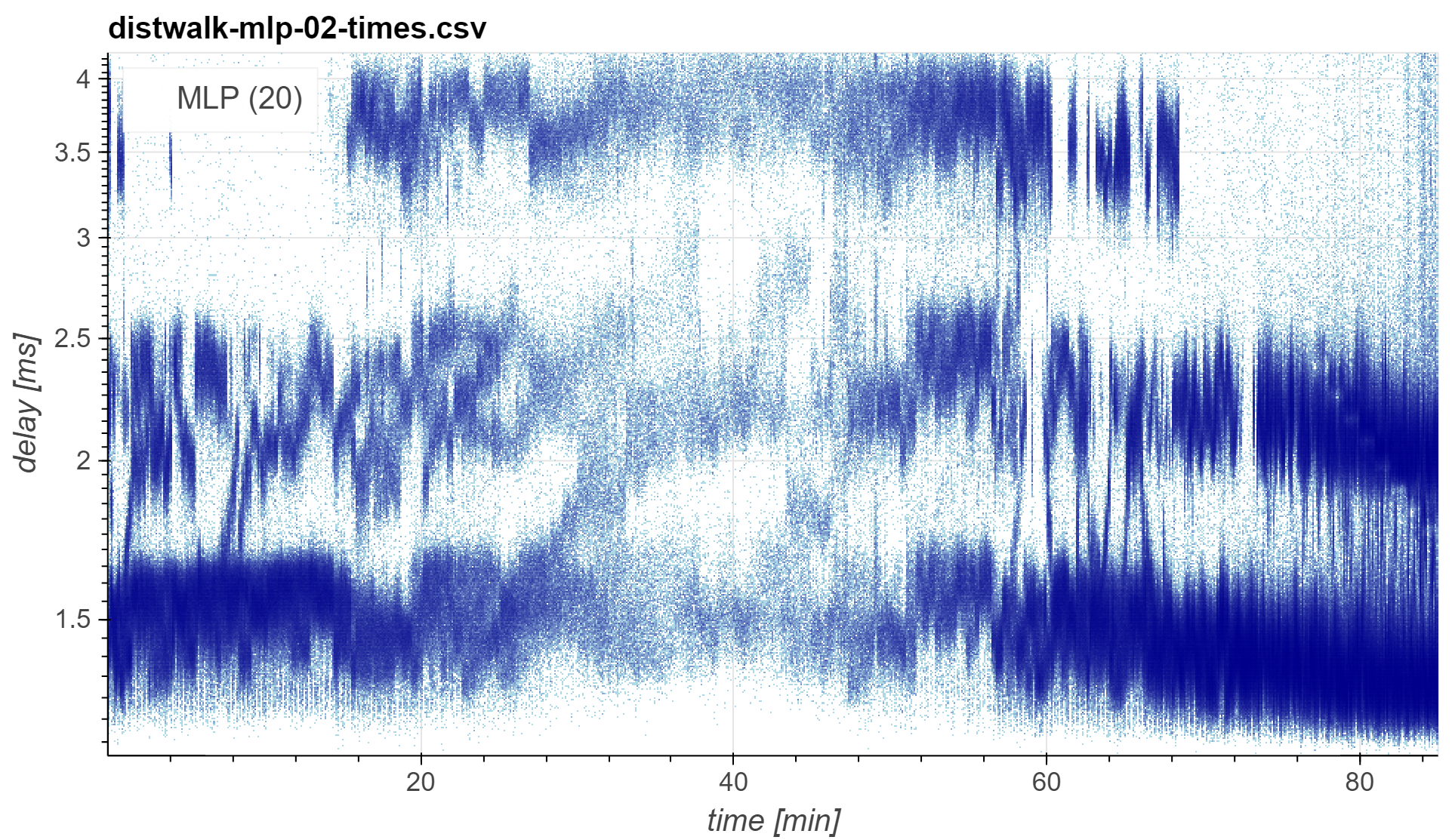}
    }
    \subcaptionbox{
        RNN-based scaling policy - CPU usage\label{fig:run-rnn}}{
        \includegraphics[keepaspectratio,width=0.48\textwidth,trim=10 0 0 60, clip]{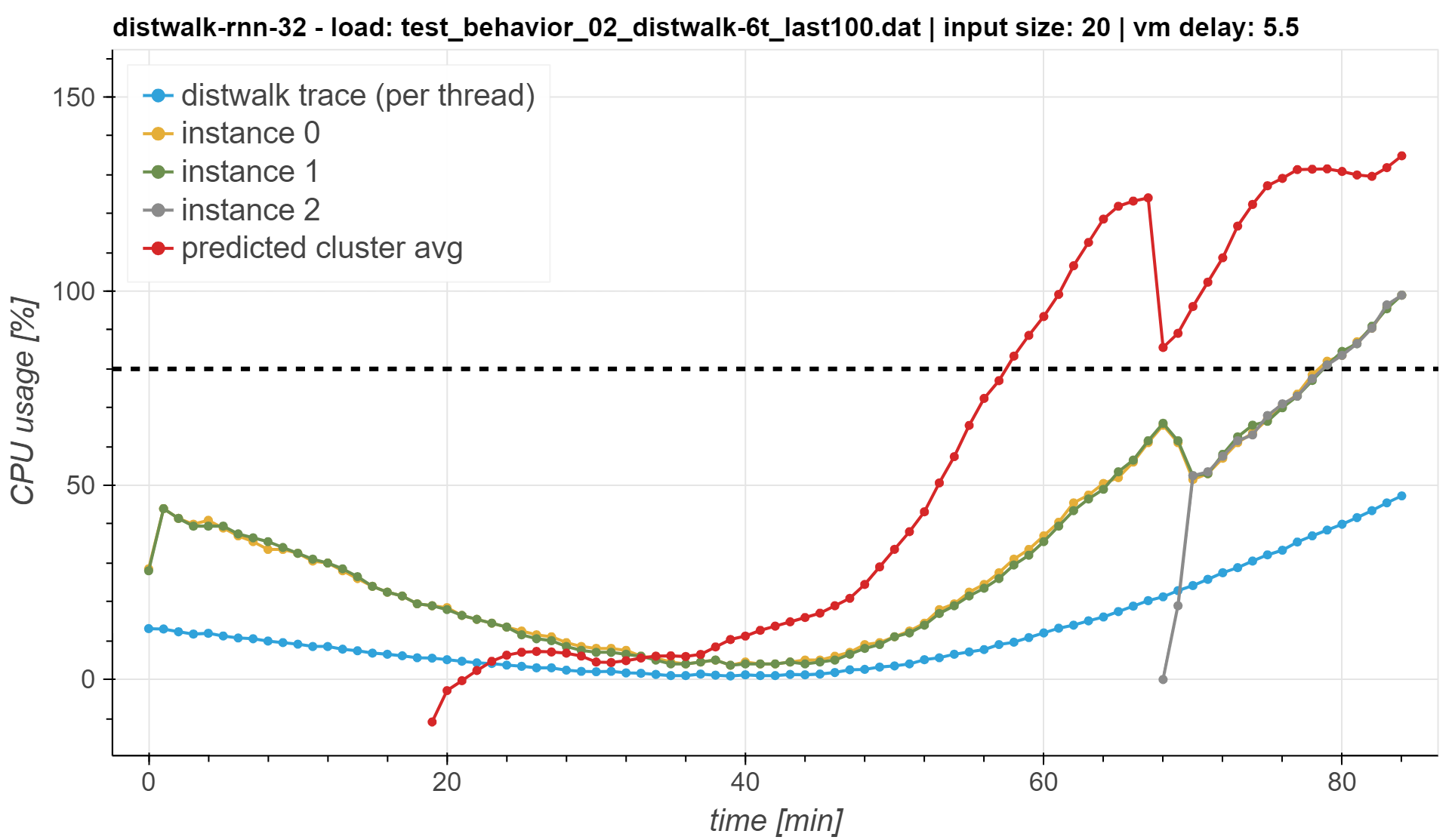}
    }
    \hfill
    \subcaptionbox{
        RNN-based scaling policy - client-side response time\label{fig:run-rnn-times}}{
        \includegraphics[keepaspectratio,width=0.48\textwidth,trim=10 0 0 60, clip]{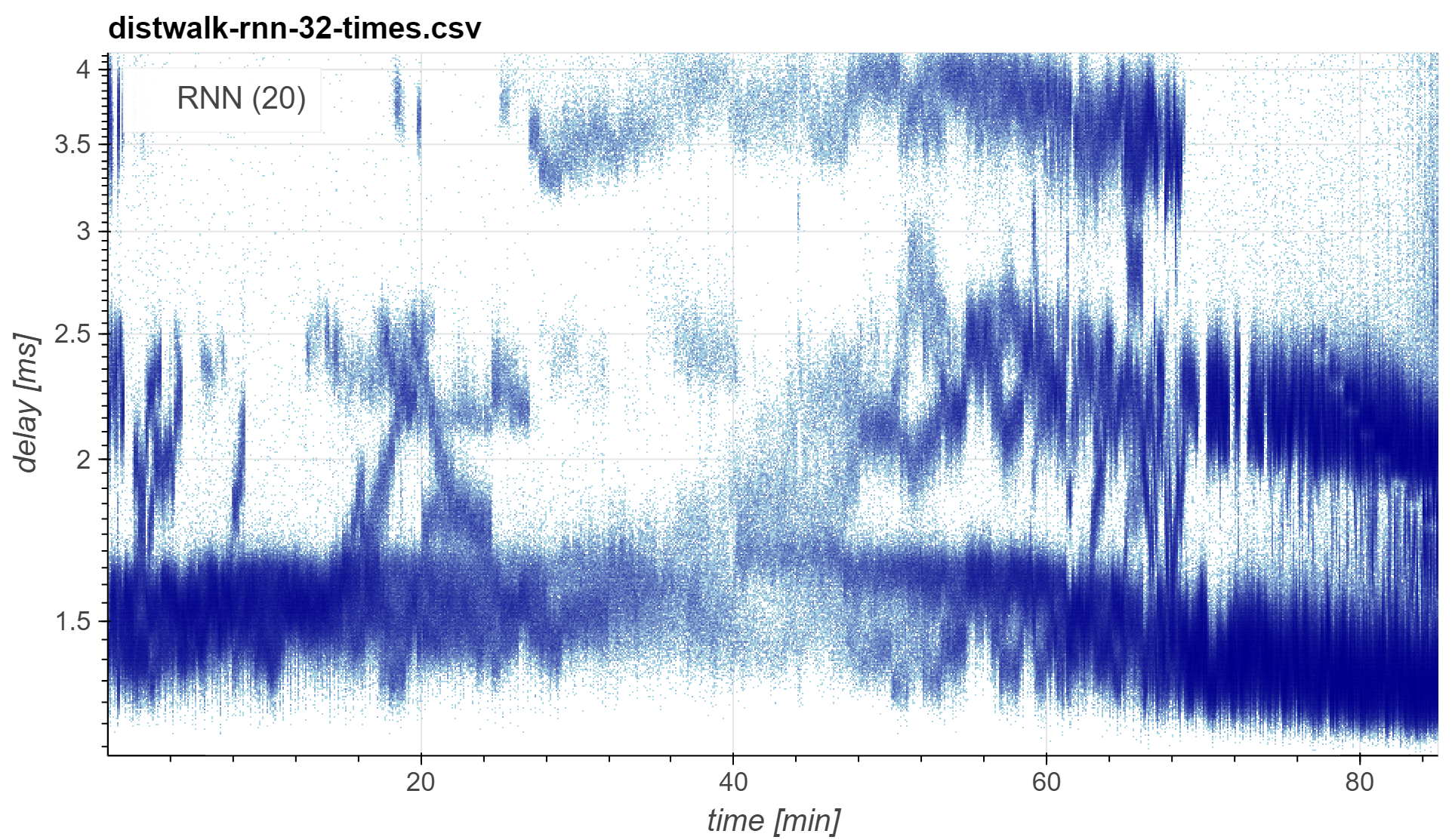}
    }
    %
    \caption{Experimental results.}
    \label{fig:results}
\end{figure*}

\Cref{fig:results} reports the obtained results from running the described
experiment, using 4 different scaling strategies, in presence of a
workload\textemdash similar to the one depicted in
\Cref{fig:load-examples}\textemdash that suddenly ramps up in a window of about
30 minutes (from time 50 to time 80 on the x-axis in the plots).
In the CPU usage plots (left-hand side), the blue curve represents the ideal
workload to be exercised on the whole cluster as input to each thread of the
distwalk client, which submits its requests to the Senlin cluster through the
Octavia load-balancer. This results in the actual per-VM workload highlighted in
the first 3 curves in each plot.
Additionally, the red curve in \Cref{fig:run-lin,fig:run-mlp,fig:run-rnn}
highlights the predicted average CPU utilization for the whole cluster, assuming
the size of the cluster to remain constant, according to the different adopted
predictive policies.

As visible, the standard threshold-based scaling strategy tuned on the average
CPU utilization crossing the $80\%$ threshold, shown in \Cref{fig:run-static},
fails to scale-up the cluster on time, as we need 3 consecutive violations of
the threshold in order to start the scale-out operation. Additionally, the
start-time of the new VM takes as much as 6 minutes, therefore the scale-out
decision happens approximately at minute 73, albeit the new VM starts serving
requests only approximately at minute 83, after the cluster has been overloaded
for about 8 minutes at full CPU utilization. Indeed, observing the obtained
client-side response times in \Cref{fig:run-static-times}, we can see that this
resulted in heavy degradation of the performance.
The other 3 plots correspond to the use of a predictive policy, with 3 different
predictors.
The simple LR-based prediction policy shown in \Cref{fig:run-lin}, performs a
better job at predicting the growth of the CPU utilization on time. Indeed, it
activates the scale-out earlier, approximately at minute 65. The anticipated
scaling has a clear effect on the observed client-side response times
(\Cref{fig:run-lin-times}), which stay all below 25 ms.
In \Cref{fig:run-mlp,fig:run-rnn}, we can see the effects of using a simple MLP
and RNN, respectively, as predictors. In these cases, we can make similar
observations about the benefits of the anticipated scaling, as in the LR-based
case. However, with respect to the latter one, the neural networks-based
approaches manage to capture the non-linear behavior of the input ramp, which
exhibits a clear upwards curvature. Therefore, they scale even slightly earlier,
approximately at minutes 60.
For instance, the MLP-based policy results in slightly better response times in
the worst case during the considered time-frame. However, overall, all
considered predictive solutions perform similarly from the view-point of the
client-side performance. For completeness, \Cref{tab:percentiles} reports the
average and various percentile values of the response-times observed by the
client during the time window shown in the previous pictures.
Additionally, \Cref{tab:overhead} reports the overhead imposed by the proposed
forecasting component at each activation that, in our experiments, happened once
per minute. Notice that the \textit{total} time includes interacting with
Monasca APIs to fetch input data.
Notice that in all experiments, even though the scaling policies expand the
cluster, reducing the per-VM workload, the input trace keeps growing, requiring
additional scale-out actions, which are inhibited due to our 20 minutes cooldown
period. This is among the engineering issues to be addressed in future
extensions of the proposed architecture.

\begin{table}[tp]
    \centering
    \caption{
        Descriptive statistics of the client-side response times observed
        during the experiments.
    }
    \begin{tabular}{rrrrr}
        \toprule
        {}         & Static & LR   & MLP  & RNN  \\
        \midrule
        avg (ms)   & 43.67  & 2.02 & 2.07 & 2.16 \\
        p90 (ms)   & 166.75 & 2.70 & 2.84 & 2.76 \\
        p95 (ms)   & 329.25 & 3.43 & 3.77 & 3.61 \\
        p99 (ms)   & 577.29 & 3.90 & 4.07 & 4.01 \\
        p99.5 (ms) & 633.81 & 4.23 & 4.20 & 4.12 \\
        p99.9 (ms) & 731.63 & 8.36 & 7.17 & 9.97 \\
        \bottomrule
    \end{tabular}
    \label{tab:percentiles}
\end{table}

\begin{table}[t]
    \centering
    \caption{
        Average overhead imposed by the proposed forecasting component.
    }
    \begin{tabular}{rrrr}
        \toprule
        {}                    & LR     & MLP    & RNN    \\
        \midrule
        forecasting time (ms) & 64.89  & 61.12  & 77.85  \\
        total time (ms)       & 285.34 & 224.66 & 308.13 \\
        \bottomrule
    \end{tabular}
    \label{tab:overhead}
\end{table}

%% file: conclusions.tex
\section{Conclusions and Future Work}

In this paper, an architecture for predictive elasticity in cloud
infrastructures has been presented. The approach, prototyped in
OpenStack leveraging on the Monasca API, allows for the automatic
creation of predictive metrics that are continuously updated
reflecting the expected evolution of the system state in a near
future. These metrics can be seamlessly combined with the regular
instantaneous ones already available through the standard OpenStack
monitoring infrastructure, to build cluster operation and management
policies that go beyond purely reactive strategies. As a case-study,
we presented an experimentation exploiting the presented architecture
for realizing a predictive elasticity controller via Senlin, that
applies autoscaling decisions trying to anticipate possible workload
surges that might not easily be handled through classical
threshold-based autoscaling rules. The proposed approach is
particularly useful for services with non-negligible instance spawning
times, as commonplace in production services in which creating new VMs
may require tens of minutes.

In the future, we plan to integrate our metric forecasting component
better within the OpenStack eco-system, introducing a number of standard
metric forecasting predictors, either univariate and multi-variate, which
can easily be deployed through the OpenStack command-line interface.

Regarding the orchestration logics, we plan to apply additional AI
techniques to experiment with alternate scaling policies (rather than
threshold-based ones), like those based on reinforcement learning, for
example.

Regarding the crucial problem of training and re-training the model, we
plan to provide a means for periodic re-training of the model based on
updated data already available in Monasca, but we also plan to introduce
a method for detection of concept drift, triggering a re-training of the
model whenever the one in use starts exhibiting a drop in accuracy.

Finally, we plan to validate at a deeper level our architecture by considering
data sets from real production workloads, and deal with scalability issues due
to the need for considering massive deployments where one might have several
predictive elasticity loops to act on several services. In this regard, we plan
to leverage on the scalable analytics processing architecture provided by Monasca
through its integrated Storm component.